\documentclass[10pt,preprint]{emulateapj}
\newcommand*{\teff}{$T_{\rm eff}$}
\newcommand*{\logg}{$\log~g$}
\newcommand*{\feh}{[Fe/H]}

\newcommand*{\cfe}{[C/Fe]}

\newcommand*{\msun}{$M_\odot$}

\newcommand*{\z}{$|Z|$}

\usepackage{epsfig}
\usepackage{apjfonts}
\usepackage{longtable}
\usepackage{graphics}
\usepackage{natbib}
\shorttitle{CARBON-ENHANCED METAL-POOR STARS IN SDSS/SEGUE II.}
\shortauthors{Lee et al.}
\slugcomment{Submitted on October 10, 2013}

\begin{document}

\title{CARBON-ENHANCED METAL-POOR STARS IN SDSS/SEGUE. II. \\
COMPARISON OF CEMP STAR FREQUENCIES WITH BINARY POPULATION SYNTHESIS MODELS}

\author{Young Sun Lee\affil{1}, Takuma Suda\affil{2}, 
        Timothy C. Beers\affil{3,4}, and Sara Lucatello\affil{5} \\
%        and Constance M. Rockosi\affil{6} \\
\scriptsize{
\textup{
\affil{1}{$^1$Department of Astronomy, New Mexico State University,
                 Las Cruces, NM, 88003, USA; yslee@nmsu.edu} \\
\affil{2}{$^2$National Astronomical Observatory of Japan, Osawa 2-21-1, Mitaka, Tokyo 181-8588, Japan} \\
\affil{3}{$^3$National Optical Astronomy Observatory, Tucson, AZ 85719, USA} \\
\affil{4}{$^4$Joint Institute for Nuclear Astrophysics (JINA), Michigan State 
                 University, East Lansing, MI 48824, USA} \\
\affil{5}{$^{5}$INAF--Osservatorio Astronomico di Padova, vicolo 
                 dell'Osservatorio 5, I-35122 Padova, Italy}}}}
%\affil{6}{$^6$UCO/Lick Observatory, Department of Astronomy and Astrophysics,
%                 University of California, Santa Cruz, CA 95064, USA}}}} 

\begin{abstract}

We present a comparison of the frequencies of carbon-enhanced 
metal-poor (CEMP) giant and main-sequence turnoff stars, selected from the 
Sloan Digital Sky Survey (SDSS) and the Sloan Extension for Galactic 
Understanding and Exploration (SEGUE), with predictions from 
asymptotic giant-branch (AGB) mass-transfer models. We consider two 
initial mass functions (IMFs)---a Salpeter IMF, and a mass function 
with a characteristic mass of 10 \msun. Previous observations 
suggest that the carbon abundances of red giants are altered during red-giant branch 
evolution due to mixing of their convective outer layers, resulting in a 
reduction of the observed carbon-abundance ratios. Thus, in order to derive 
more accurate estimates of CEMP frequencies for stars in the Milky Way, 
it is preferable to make use of SDSS/SEGUE main-sequence turnoff stars, 
which are not expected to experience significant dilution. However, because 
of the difficulty of identifying moderately carbon-enhanced stars 
($+0.7 <$ \cfe\ $<+1.5$) among warm, metal-poor turnoff stars, owing to their 
much weaker CH $G$-bands, we derive a correction function to compensate for 
the resulting undercounts of CEMP stars. These comparisons indicate good 
agreement between the observed CEMP frequencies for stars with [Fe/H] $>
-1.5$ and a Salpeter IMF, but not with an IMF having a higher characteristic mass.
Thus, while the adopted AGB model works well for low-mass progenitor stars, it does
not do so for high-mass progenitors. Our results imply that the IMF shifted
from high- to low-mass dominated in the early history of the Milky
Way, which appears to have occurred at a ``chemical time'' between 
[Fe/H] $= -2.5$ and [Fe/H] $= -1.5$. The corrected CEMP frequency for the
turnoff stars with [Fe/H] $< -3.0$ is much higher than the AGB model
prediction from the high-mass IMF, supporting the previous assertion
that one or more additional mechanisms, not associated with AGB stars, are
required for the production of carbon-rich material below [Fe/H] $= -3.0$.

\end{abstract}


\section{Introduction}

Numerous spectroscopic studies of metal-poor ([Fe/H] $< -1.0$)
candidates identified by the HK survey (Beers et al. 1985, 1992) and the
Hamburg/ESO Survey (HES; Wisotzki et al. 1996; Christlieb et al. 2001, 2008; 
Christlieb 2003) have revealed that the frequency of
carbon-enhanced stars increases strongly with decreasing [Fe/H]. These
stars, now known as carbon-enhanced metal-poor (CEMP) stars, were
originally defined as stars with metallicity [Fe/H] $\leq -1.0$ and
carbon-to-iron ratios [C/Fe] $\geq +1.0$ (Beers \& Christlieb 2005).
\footnote[6]{Different criteria, such as [C/Fe] $> +0.5$ and [C/Fe] $>
+0.7$ are used by a number of studies as well; most recent work assumes
[C/Fe] $> +0.7$.} Generally, the frequency
of C-rich stars increases from a few percent at higher metallicity to on
the order of 20\% for [Fe/H] $< -2.0$, 30\% for [Fe/H] $< -3.0$, 40\%
for [Fe/H] $< -3.5$, and 75\% for [Fe/H] $< -4.0$ (Beers et al. 1992;
Norris et al. 1997; Rossi et al. 1999; Beers \& Christlieb 2005; Cohen
et al. 2005; Marsteller et al. 2005; Rossi et al. 2005; Frebel et al.
2006; Lucatello et al. 2006; Norris et al. 2007; Carollo et al. 2012;
Norris et al. 2013; Spite et al. 2013; Yong et al. 2013). This
increasing trend of CEMP-star frequency with declining [Fe/H] is again
confirmed from the many thousands of CEMP stars found among the several
hundred thousand stars with available spectra from 
the Sloan Digital Sky Survey (SDSS; Fukugita et al. 1996; Gunn et al. 1998, 2006; 
York et al. 2000; Stoughton et al. 2002; Abazajian et al. 2003, 2004, 2005, 2009; Pier et
al. 2003; Adelman-McCarthy et al. 2006, 2007, 2008; Aihara et al. 2011;
Ahn et al. 2012) and the Sloan Extension for Galactic Understanding and
Exploration (SEGUE-1; Yanny et al. 2009), and SEGUE-2 (C. Rockosi et
al., in preparation) as described by Lee et al. (2013). 

There exist a number of subclasses within the CEMP classification, as
originally defined by Beers \& Christlieb (2005), which may provide
direct clues to the nature of their likely progenitors. Stars in the
CEMP-$s$ subclass exhibit over-abundances of $s$(low)-process elements
such as Ba and Sr, the CEMP-$r$ subclass includes stars with enhanced
$r$(apid)-process elements such as Eu, and the CEMP-$r/s$ stars exhibit
elemental abundance patterns associated with both the $r$-process and
the $s$-process. The CEMP-no subclass exhibits no over-abundances of
the neutron-capture elements.

The CEMP-$s$ (and CEMP-$r/s$) subclasses of CEMP stars are the most
commonly found to date; high-resolution spectroscopic studies show that
around 80\% of the CEMP stars are categorized as CEMP-$s$ (or
CEMP-$r/s$) (Aoki et al. 2007, 2008). The favored mechanism for the
production of the high [C/Fe] ratios found for CEMP-$s$ (CEMP-$r/s$)
stars is mass transfer of carbon-enhanced material from the envelope of
a now-defunct asymptotic giant-branch (AGB) star to its (presently
observed) binary companion (Suda et al. 2004; Herwig 2005; Komiya et al.
2007; Sneden et al. 2008; Masseron et al. 2010; Bisterzo et al. 2011,
2012). Observational evidence now exists to suggest that the CEMP-$r/s$
stars (and other $r$-process-element rich stars) were enhanced in
$r$-process elements in their natal gas clouds by previous generations
of supernovae (SNe), and did not require a contribution of $r$-process
elements from a binary companion (see Hansen et al. 2013).

The limited amount of long-term radial-velocity monitoring available for
CEMP stars indicates variations for almost all of the CEMP-$s$ stars,
confirming their binary status (Lucatello et al. 2005a). In addition,
the CEMP-$s$ stars are mostly, though not exclusively (e.g., Norris et
al. 2013 and references therein), found among metal-poor stars with
[Fe/H] $> -3.0$. On the other hand, CEMP-no stars are found most
commonly among the extremely metal-poor (EMP) stars, with [Fe/H] $<
-3.0$ (Aoki et al. 2007; Norris et al. 2013). Existing radial-velocity
monitoring of these objects indicates that they are found in binary
systems no more frequently than other metal-poor stars (T. Hansen et
al., in preparation). Norris et al. (2013) found no CEMP-$s$ stars among
18 CEMP stars with [C/Fe] $\geq +0.7$ and [Fe/H] $< -3.1$, as well as no
discernible variations of their radial velocities.

Although there is general consensus on the origin of CEMP-$s$ stars, the
likely progenitor or progenitors of the CEMP-no stars are still under
discussion. Suggested models include massive, rapidly rotating, mega
metal-poor (MMP; [Fe/H] $< -6.0$) stars, which produce large amounts of
C, N, and O due to distinctive internal burning and mixing episodes
(Meynet et al. 2006, 2010; Chiappini 2013), and faint (low-energy) SNe
associated with the first generations of stars, which experience
extensive mixing and fallback during their explosions, and eject large
amounts of C and O, but not heavier metals (Umeda \& Nomoto 2003, 2005;
Tominaga et al. 2007, 2013; Ito et al. 2009, 2013; Nomoto et al. 2013).
Nevertheless, the origin of the CEMP-no star phenomenon is yet to be
fully resolved (see Norris et al. 2013, which summarizes other possible
progenitors of the C-rich stars).

Previous authors have attempted to understand the large fractions of
CEMP stars at low metallicity, as well as the different subclasses of
the CEMP stars, by invoking AGB models with different masses.
Furthermore, there have been several efforts to constrain the form of
the early initial mass function (IMF) by reproducing the observed
frequencies of CEMP stars, as well as the number ratios of the different
CEMP subclasses. Abia et al. (2001), for example, claimed that the large
number of carbon-enhanced stars found among stars of very low
metallicity could be accounted for if the IMF in the early history of
the Galaxy was dominated by higher mass stars. Lucatello et al. (2005b)
and Komiya et al. (2007) utilized population-synthesis models with an
IMF biased towards massive stars to compare with the fractions of
observed CEMP stars, and concluded that an IMF comprising a larger
number of intermediate- to high-mass stars could reproduce the larger
fraction of the CEMP stars among metal-poor stars ([Fe/H] $< -2.5$)
better than the present-day (Salpeter) IMF. Recently, Suda et al. (2013)
made use of the number ratios of CEMP/EMP, CEMP-no/CEMP, and NEMP/CEMP
giant stars (where NEMP stands for nitrogen-enhanced metal-poor) from
the Stellar Abundances for Galactic Archeology (SAGA; Suda et al. 2008)
database to constrain the parameters in their binary population-synthesis
model. They considered several IMFs, and proposed that the IMF changed
from high-mass dominated in the early Galaxy to low-mass ($M < 0.8$
\msun) dominated at present, and that this transition occurred around a
metallicity of [Fe/H] $\sim -2.0$. 

The above studies carried out comparisons of the CEMP fractions derived
from small samples of stars comprising mostly giants. However, 
observational evidence indicates that the C-rich material at the
surface of a giant could be easily depleted by extra mixing of
CNO-processed material from its interior during the so-called first
dredge-up episode (Spite et al. 2005, 2006; Lucatello et al. 2006; Aoki
et al. 2007). It is also known that more luminous red giant-branch (RGB)
stars are more affected by such mixing (Spite et al. 2005, 2006).
Therefore, if such mixing does occur, the overall CEMP frequencies as
estimated from giants are expected to be a lower limit.  

In order to avoid this complication, the best way forward would appear
to be comparing model predictions with the observed CEMP frequencies based
on unevolved stars, such as dwarfs or main-sequence turnoff stars. The
current CEMP stars that have been studied with high-resolution
spectroscopy are mostly giants (for which it is simpler to obtain high
S/N spectra, due to their relative brightness and moderate temperatures,
which allows for lines of interest to be measured with less
uncertainty). The numbers of observed dwarf and turnoff stars with similar
observations are in any case too small to derive statistically
meaningful results for different subclasses of CEMP stars. 

In this study, we make use of stars with available carbon-to-iron ratios
(\cfe) and [Fe/H], based on medium-resolution ($R \sim 2000$)
spectroscopy obtained during the course of the SDSS, SEGUE-1, and SEGUE-2, 
in order to derive accurate frequencies of CEMP
stars among giants and turnoff stars as a function of [Fe/H]. The
derived CEMP frequencies are then compared with the predictions from AGB
binary-synthesis models that employ the two different IMFs explored by
Suda et al. (2013). The results of these comparisons should provide more
stringent constraints on the IMF of the Milky Way, and clues to the
existence of progenitors {\it other than AGB stars} that are capable of
producing large amounts of carbon-enhanced material in the early
universe. 

This paper is outlined as follows. In Section 2, we describe the
selection criteria used to assemble the sample for this study. Section 3
presents and discusses results of the comparison of the CEMP frequencies
for giants and main-sequence turnoff stars with the binary-synthesis
model predictions, and describes a procedure for correcting the
anticipated undercounts of CEMP stars among warm, metal-poor turnoff
stars. Our conclusions are presented in Section 4.

\section{Carbon-enhanced SDSS/SEGUE Stars}

The SDSS, SEGUE-1, and SEGUE-2 surveys have produced an unprecedented
number of high-quality medium-resolution stellar spectra, covering stars
in various evolutionary stages, and spanning a wide range of metallicity
($-$4.0 $<$ [Fe/H] $< +0.5$). A total of about 600,000 stars are
potentially suitable for examination of the properties of the Milky
Way's stellar populations. The resolving power of the spectra is
$R \sim 2000$, over the wavelength range 3820--9100\,\AA. Below we simply
refer to these stars (spectra) as SDSS/SEGUE stars (spectra). Accurate
estimates of the atmospheric parameters for most of the SDSS/SEGUE stars
are derived using the latest version of the SEGUE Stellar Parameter
Pipeline (SSPP; Lee et al. 2008a, 2008b, 2011; Allende Prieto et al.
2008; Smolinski et al. 2011). The typical external errors obtained by
the SSPP are 180 K for \teff, 0.24 dex for \logg, and 0.23 dex for \feh,
respectively (Smolinski et al. 2011). In addition, estimates of the
carbonicity, \cfe, is obtained following the prescription of Lee et al.
(2013), for stars with $4400 \leq$ \teff\ $\leq$ 6700 K, where accurate 
[C/Fe] can be determined. As reported by Lee et al., uncertainties in the 
determination of [C/Fe] are smaller than 0.35 dex for SDSS/SEGUE 
spectra with S/N $\geq 15$ \AA$^{-1}$. 

%Figure 1
\begin{figure}
\centering
\plotone{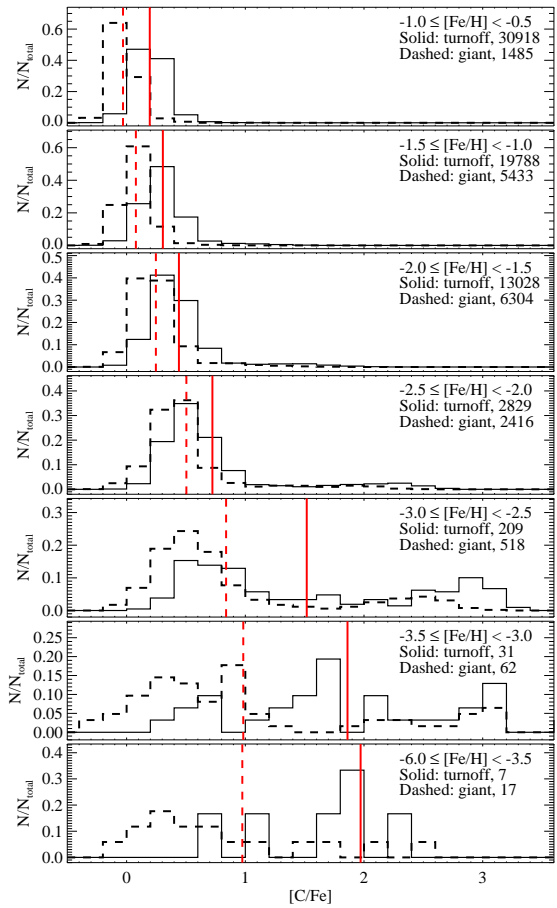}
\caption{Distribution of [C/Fe] for SDSS/SEGUE stars over different
ranges of [Fe/H], decreasing from the upper to lower panels.
Main-sequence turnoff stars are shown as solid histograms; giants are
shown as dashed histograms. The metallicity range in each panel
is indicated by the legends, along with the total number of
stars shown in the histograms. The solid and dashed red vertical lines
indicate the mean values of [C/Fe] for the turnoff stars and giants,
respectively. On average, the turnoff stars exhibit higher carbon
enhancement than the giants.}
\label{fig:cfedist}
\end{figure}

In order to derive reliable frequencies of the CEMP stars among the field 
stellar populations, we follow the selection criteria of Lee et al.
(2013). Briefly, we first exclude all stars located in the directions of
known open and globular clusters. For stars that were observed more than
once, we keep only the parameters derived from the highest S/N spectrum.
We then restrict the sample to stars with spectra having S/N $\geq 20$
\AA$^{-1}$, effective temperatures in the range 4400~K $\leq$ \teff\
$\leq$ 6700~K, and metallicities in the range $-$4.0 $\leq$ \feh\ $\leq
+0.5$, so that our estimates of \cfe\ are as reliable as possible. 

We then visually inspect each spectrum with [Fe/H] $\leq -2.0$, in order
to reject spectra such as cool white dwarfs, or those with emission-line
features in the cores of their Ca\,{\sc ii} lines, or other spectral
defects that could lead to spurious determinations of metallicity by the
SSPP. Additionally, we visually examine the spectra for all stars with \cfe\
$\geq +0.7$, and exclude stars with poor estimates of \feh\ and/or \cfe.
Following Lee et al., for the purpose of deriving the CEMP frequencies
we do not include stars with [C/Fe] $\ge +0.7$ and 
indication of upper limit estimate, and
consider these stars to have unknown carbon status. By application of
the above procedures, we are left with a sample of about 247,350 stars.

For the purpose of our analysis, we consider stars with 4400 K $\leq$
\teff\ $\leq$ 5600 K and 1.0 $\leq$ \logg\ $<$ 3.2 as giants, and stars
with 5600 K $\leq$ \teff\ $\leq$ 6600 K and 3.2 $\leq$ \logg\ $<$ 4.5 as
main-sequence turnoff stars. The distance to each star is estimated 
following the prescriptions of Beers et al. (2000, 2012), which obtains
photometric distance estimates with errors on the order of 10\%--20\%.

Note that we have added stars to our sample from Table 1 of Yong et al.
(2013), and one object from Caffau et al. (2011), with determinations
based on high-resolution spectroscopic analyses, in order to increase
the number of stars with [Fe/H] $<-3.0$. This results in better number
statistics for the calculation of the CEMP frequencies in the extremely
and ultra metal-poor regime.

%Figure 2
\begin{figure*}
\centering
\plotone{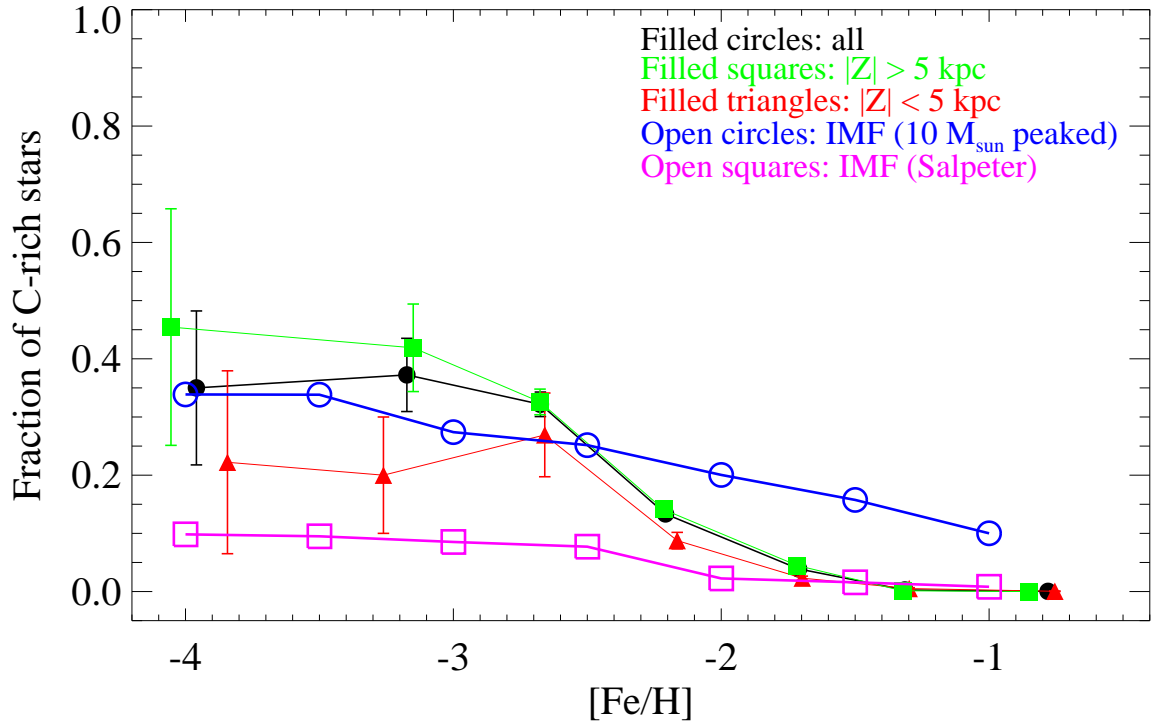}
\caption{Differential frequencies of C-rich giants (defined as stars with 
4400 K $\leq$ \teff\ $\leq$ 5600 K and 1.0 $\leq$ \logg\ $<$ 3.2) as a
function of [Fe/H]. Bin sizes of 0.5 dex in [Fe/H] are used for stars
with [Fe/H] $> -3.5$; a single bin for stars with [Fe/H] $< -3.5$ is
used. Filled circles show the derived frequencies of the CEMP stars for
all giants, while the filled squares represent the subsample of CEMP
stars for giants located more than 5 kpc from the Galactic plane, \z\
$>$ 5 kpc. The filled triangles indicate stars located closer to the
plane, with \z\ $<$ 5 kpc. The prediction of the expected CEMP
frequencies as a function of [Fe/H], based on an AGB nucleosynthesis
model with IMF peaked at 10 \msun, is indicated by the open circles,
while the open squares denote the prediction assuming a Salpeter IMF.
Error bars are based on Poisson statistics.}
\label{fig:cempcom}
\end{figure*}

\section{Results and Discussion}

\subsection{Differences in Average [C/Fe] Between Giants and Turnoff Stars}

Figure \ref{fig:cfedist} shows the distribution of [C/Fe] for SDSS/SEGUE
stars in various bins of metallicity, decreasing from the upper to lower
panels. Main-sequence turnoff stars are shown as solid histograms;
giants are shown as dashed histograms. Inspection of this figure reveals
that the overall distribution of [C/Fe] gradually shifts (for both
turnoff stars and giants) to higher [C/Fe] with decreasing [Fe/H], with
a tail extending toward higher [C/Fe] appearing as the metallicity
decreases. As the metallicity decreases below [Fe/H] $= -3.0$, this
trend continues for the turnoff stars, but it is not as evident for the
giants.

Another interesting feature seen in Figure \ref{fig:cfedist} is that,
for [Fe/H] $< -3.0$, the turnoff stars are distributed over a wide range
of [C/Fe], whereas the giants are mostly concentrated in the region of
[C/Fe] $< +1.0$. The red vertical lines in Figure \ref{fig:cfedist}
indicate the mean values of [C/Fe] for the turnoff stars (solid line)
and giants (dashed line), respectively. On average, the giants appear to
exhibit lower carbonicity (by about 0.2 dex) than the turnoff stars,
down to [Fe/H] = $-2.5$. The mean value of [C/Fe] appears to increase
with decreasing metallicity, as also found by Carollo et al. (2012,
their Figure 11). They reported that the degree of carbon enhancement
significantly increased from [C/Fe] $\sim +1.0$ at [Fe/H] = $-1.5$ to
[C/Fe] $\sim +1.7$ at [Fe/H] = $-2.7$, somewhat higher than our values
(it should be noted that a different sample of stars, as well as a
different method for determination of [C/Fe], were employed by these
authors).

The difference in the distribution of [C/Fe] between the turnoff stars
and the giants may be explained by the different masses of the
convective envelopes between the two evolutionary stages. Because a
giant has a much larger convective envelope, its surface material
experiences more mixing, leading to reduction of the carbonicity. On the
other hand, the turnoff stars have shallower convective envelopes, so
that their surface abundances may not be expected to greatly change. As
a result, the overall carbon abundance for the turnoff population is
expected to be higher than that of the giant population, even if they
were born with the same initial carbon abundance. 

Bonifacio et al. (2009) also noted a difference in the mean [C/Fe]
between giants and turnoff stars of similar metallicities, finding a
difference of about 0.2 dex (giants being lower) for stars with [Fe/H]
$< -2.5$. They argued that this difference arises because the giants
suffer from extra mixing due to first dredge-up, and have their surface
carbon abundance reduced. They also suggested the stellar models
employed in the analysis could contribute to this discrepancy; they
found a smaller difference between the giants and turnoff stars when
deriving [C/Fe] from a 3D, rather than a 1D model atmosphere.

\subsection{Comparison with Model Predictions for the Frequency of
C-rich Giants}

The black filled circles in Figure \ref{fig:cempcom} represent our
derived differential frequencies for C-rich giants as a function of
\feh. From inspection of this figure, it appears that the CEMP
frequencies do not increase for [Fe/H] $<-2.5$, but rather, remain 
relatively constant, in contrast to the results of previous
studies. As discussed by Spite et al. (2005, 2006), Lucatello et al.
(2006), and Aoki et al. (2007), this may be in part due to
CN-processing, which converts carbon to nitrogen at the bottom of a
star's convection zone, and in turn reduces the carbon abundance in the
envelope at the time of first dredge-up. Whether or not a giant
experiences such mixing can be identified by measuring its
$^{12}$C/$^{13}$C or [C/N] ratios, as both will be lower for an
object that has gone through such an event. Unfortunately, these ratios are
difficult to assess from the SDSS/SEGUE spectra over the full range of
metallicities we consider. In the metallicity regime of [Fe/H] $> -1.5$, 
our derived CEMP frequencies of $\sim$1\% agree with the
previously claimed fraction of classical CH or Ba stars in the solar
neighborhood (Luck \& Bond 1991). 

Regarding the effect of the first dredge-up, it is worth mentioning the
following arguments from a theoretical point of view. According to Suda
et al. (2004), first dredge-up might not play a significant role in
decreasing the carbon abundance on the surface of a giant, based on an
0.8 \msun\ model for HE~0107-5240 (a CEMP-no star with [Fe/H] = $-5.3$;
Christlieb et al. 2002). They examined the effect of the first dredge-up
following the accretion of C- and O-rich matter onto the star (here
assuming an AGB mass-transfer scenario), and found that if its envelope
is significantly C-rich ([C/Fe] $\gg +1.0$), then after first dredge-up
the surface carbon abundance changed little. This is because the surface
carbon abundance is too large ([C/H] $\sim -1$) prior to the dredge-up
to be significantly depleted by the relatively small amount of matter in
the hydrogen-burning shell, $M <$ 0.02 \msun. Even in cases of EMP stars
for which the initial carbon abundance is small, Suda \& Fujimoto (2010)
showed that the effect of the first dredge-up is also limited, due to
the shallower convective envelopes in metal-poor (as compared to
metal-rich) stars. According to their model calculation, the change of
the CNO abundance before and after the first dredge-up was on the order
of one percent. Therefore, they did not notice a large impact on the
surface carbon abundances after first dredge-up for stars with [Fe/H] $<
-2.3$.

The blue open circles in Figure \ref{fig:cempcom} are the predicted
frequencies of CEMP giants, as a function of metallicity, from AGB
binary-synthesis models with an IMF peaked at 10 \msun, while the open
squares are the predicted frequencies from models using a Salpeter IMF,
adopted from Suda et al. (2013). In their model, they included a
mechanism referred to as ``pulsation-driven mass loss'' (Wood 2011),
which was argued to suppress the previously predicted over-production of
NEMP stars by Izzard et al. (2009) and Pols et al. (2012). It appears
that the predicted CEMP frequencies for the high-mass dominated IMF are in
relatively good agreement with the observed CEMP giant frequencies for
[Fe/H] $< -2.5$, but the model predicts too many C-rich stars above [Fe/H] =
$-2.5$. The predicted CEMP frequencies from a Salpeter IMF are in good
agreement with our derived frequencies for the metal-rich region ([Fe/H]
$> -1.5$), while the predicted CEMP frequencies are far too low for
stars with [Fe/H] $< -2.5$. 

These are similar results to those found by Suda et al. (2013), who used
the giants in the SAGA database to compare the observed CEMP frequencies
with their model predictions. One of the reasons that Suda et al. (2013)
employed giants to derive the CEMP frequency is that one can ignore
effects such as atomic diffusion, which can alter the surface abundances
of dwarfs and turnoff stars more significantly than in giants (e.g.,
Richard et al. 2002a,b; Korn et al. 2007; Lind et al. 2008). However,
our derived frequencies show a much better agreement for the Salpeter
IMF in the metallicity region [Fe/H] $> -1.5$. In the study of Suda et
al. (2013), the model-predicted frequency of the CEMP stars above [Fe/H]
= $-2.0$ was not well-constrained, most likely due to the selection
biases associated with the assembly of their sample from previous
high-resolution spectroscopic studies (which tended to emphasize the
more metal-poor and/or carbon-enhanced stars). In contrast, the good
agreement of the frequencies calculated from our considerably
less-biased SDSS/SEGUE sample with the model prediction for [Fe/H] $>
-1.5$ suggests that the AGB binary-synthesis model with a Salpeter mass
function used by Suda et al. (2013) works well, at least in this
metallicity regime.

Based on the results from the comparisons of the observed CEMP
frequencies with model predictions from the two different IMFs, we
conjecture that, for very low-metallicity ([Fe/H] $< -2.5$) stars, the
distribution of the stellar masses was dominated by rather massive stars
($\sim$10 \msun\ or higher), while for the relatively more metal-rich
stars ([Fe/H] $> -1.5$), it appears that the IMF did not much differ
from a Salpeter IMF, which is biased towards low-mass progenitor stars
($M < 0.8$ \msun). As previously claimed by Suda et al. (2013), our
results also support the idea that there must exist a shift in the IMF
from a high-mass dominated to low-mass dominated form in the early
history of the Milky Way, corresponding to a ``chemical time'' between
[Fe/H] $= -2.5$ and [Fe/H] $= -1.5$. 

By way of comparison, the binary population-synthesis model of Izzard et
al. (2009) was able to reproduce the ratio of NEMP to very metal-poor
(VMP; [Fe/H] $< -2.0$) stars (that is, C and N normal stars) {\it
without} introducing an IMF dominated by higher mass stars, but not the
high frequency of the CEMP stars. Pols et al. (2012) also argued, by
comparing the observed number ratio of NEMP to CEMP stars with their
model predictions, that they could derive a similar number ratio from a
Salpeter IMF, and ruled out an IMF peaked at 10 \msun\ claimed by Komiya
et al. (2007).

\subsection{Behavior of Derived CEMP Frequencies with Distance from the Galactic
Plane}

Another interesting result emerges when one partitions the giant sample
based on distance from the Galactic plane. The green squares in Figure
\ref{fig:cempcom} are the CEMP frequencies for giants with distances from
the Galactic mid-plane ($|Z|$) larger than 5 kpc, whereas the red
triangles represent frequencies based on those with $|Z| < 5$
kpc.\footnote[7]{A more quantitative analysis of the variation of CEMP
frequencies with $|Z|$ will be considered in an upcoming paper of this series.} The
figure clearly indicates that the more distant halo giants exhibit 
higher frequencies of C-rich stars, while the stars closer to the Galactic
plane tend to have lower frequencies of C-rich stars. This same trend
with vertical distance was hinted at (due to small number statistics) in
Frebel et al. (2006), and strongly confirmed in the much larger sample
of SDSS/SEGUE calibration stars by Carollo et al. (2012).  Carollo et
al. argued that this result was likely due to the fact that
the outer-halo population has about twice the frequency of CEMP stars, at
a given low metallicity, as the inner-halo population.

A few possible reasons for the observed differences in the CEMP
frequencies between the two spatial regions might be suggested within
the context of the AGB model predictions. First, the progenitors of the
inner-halo population (which dominates for $|Z| <$ 5 kpc) and the
outer-halo population (which, at $|Z| >$ 5 kpc, includes more outer-halo
stars) might have formed their stars at different times, with different
IMFs. Because the outer-halo population has more CEMP stars than the
model prediction for \feh\ $<-2.5$, it is possible that the outer-halo
population might have had an IMF with more intermediate-mass stars than
considered by the model. On the other hand, since the CEMP frequencies of
the inner-halo stars are lower than the model estimate, the inner-halo
population might have had an IMF with less intermediate-mass stars than
the proposed IMF. Related ideas are discussed by Tumlinson et al. (2007). 

Another possibility is that Suda et al. (2013) assumed that all CEMP
stars, including CEMP-no objects, formed from the AGB binary scenario.
If there were to exist other channels of carbon production at [Fe/H]
$<-2.5$, such as faint SNe or rapidly rotating massive stars (producing
CEMP-no stars in the subsequent generation), as suggested by several
studies, we then might expect larger frequencies of CEMP stars than the
AGB binary-synthesis model prediction (as seen in Figure 
\ref{fig:cempcom}), even if the carbon dilution of 
the giants due to extra mixing is taken into account. A more detailed 
discussion of this is provided below.

In any event, the frequency difference we find can be understood (as
argued by Carollo et al. 2012) as the result of a change in the dominant
population with distance above the plane, from the inner-halo population to the
outer-halo population. Carollo et al. further argued that the inner halo is
dominated by stars with modest carbon enhancement ([C/Fe] $\sim +0.5$),
while the outer halo has a greater portion of stars with large carbon
enhancements ([C/Fe] $\sim +2.0$), although considerable overlap still
exists. They interpreted these results, as well as the increase in the
global frequency of CEMP stars with distance from the Galactic plane, as
evidence for the possible presence of additional astrophysical sources
of carbon, beyond AGB production alone, associated with the progenitors
of the outer-halo stars.

It is difficult to separate CEMP-$s$ and CEMP-no stars from our
medium-resolution SDSS/SEGUE spectra, but determination of the ratio of
CEMP-$s$/CEMP-no for stars in the inner- and outer-halo populations,
based on high-resolution spectroscopy, will provide not only very strong
constraints on the binary-synthesis model, but clues to the origin of
the different CEMP frequencies between the inner- and outer-halo
populations.

\subsection{Correcting the CEMP Frequencies for Main-Sequence Turnoff Stars}

Above we have compared the derived CEMP frequencies from our sample of
giants with the predicted CEMP frequencies from AGB binary-synthesis
models by Suda et al. (2013). However, as previously mentioned,
observations suggest that a giant can suffer from dilution of the
carbon-rich material in its envelope by mixing during the first
dredge-up event, resulting in a lower overall carbon abundance and (as a
population) lower frequencies of CEMP stars. Absent such dilution, we
would expect that the actual frequencies of CEMP stars among
giants would be higher than shown in Figure \ref{fig:cempcom}.

%Figure 3
\begin{figure}
\centering
\plotone{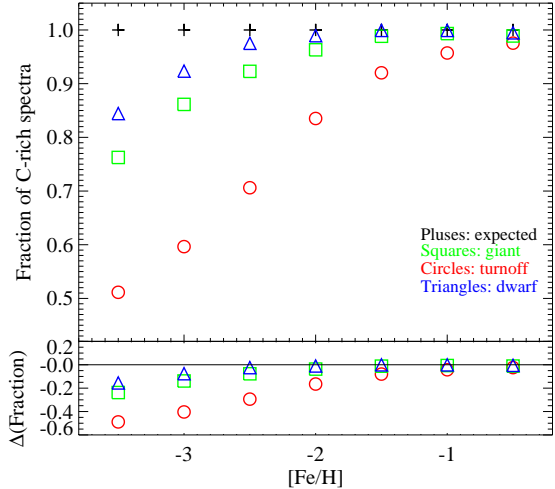}
\caption{Fractions of noise-added synthetic spectra with \cfe\ $< +0.7$
estimated by the SSPP among model spectra with mild carbon enhancements, in the range
$+0.75 \leq$ [C/Fe] $\leq +1.25$. The plus signs are the actual expected
fractions (set for convenience to unity). The green squares indicate the
calculated fraction of CEMP stars for giants, defined as models with
parameters 4500~K $\leq$ \teff\ $\leq$ 5500~K and 1.0 $\leq$ \logg\
$\leq 3.0$. The red circles are the calculated frequencies for turnoff
stars, defined as models satisfying with 5750~K $\leq$ \teff\ $\leq$
6500~K and 3.5 $\leq$ \logg\ $\leq 4.0$. The blue triangles represent
the frequencies for main-sequence dwarfs, defined to have  
4500~K $\leq$ \teff\ $\leq$ 5500~K and 4.5 $\leq$ \logg\ $\leq 5.0$. 
The bottom panel shows the distribution of residuals in the fractions
for the three populations.}
\label{fig:corfun}
\end{figure}

Stars near the main-sequence turnoff region do not experience dredge-up
episodes; rather, they preserve unpolluted material on their surfaces.
It might be possible that their surface abundances could be affected by
atomic diffusion, but because the impact on the carbon abundance is not
well known, we do not take this concern into consideration in this
study. Thus, we expect that one could obtain a more valid estimate of
the frequencies of CEMP stars in a given population by making this
evaluation using main-sequence turnoff stars. Since turnoff stars evolve
quickly into giants during their evolution, we might expect that the
frequencies of CEMP stars inferred from stars near the main-sequence
turnoff should be the same as for giants that have not yet mixed
carbon-depleted material into their envelopes. Therefore, it is
desirable to compare the predictions from the models to the frequencies
of CEMP stars derived from the turnoff stars. 

However, additional complications exist. The stars located near the
main-sequence turnoff are relatively warmer than the red giants, and as
a result, for a given carbon abundance, the molecular CH $G$-band feature
becomes significantly weaker. To make matters more difficult, at low
metallicity (assuming carbon is not enriched) a star's CH $G$-band will
also become lower in strength. Even with high-resolution spectroscopy
Aoki et al. (2013) noted that, although they were able to detect the CH
$G$-band for a star with \cfe\ $\geq +1.5$ and [Fe/H] $\sim -3.0$ at
\teff\ $\sim 6000$ K, they failed to measure the CH $G$-band for
[C/Fe] $<+1.5$ in their sample of very metal-poor stars. These effects
become even more prevalent for medium-resolution spectra, hence the
calculated CEMP frequencies obtained from the turnoff stars may also
be lower than the actual values.

%Figure 4
\begin{figure*}
\centering
\plotone{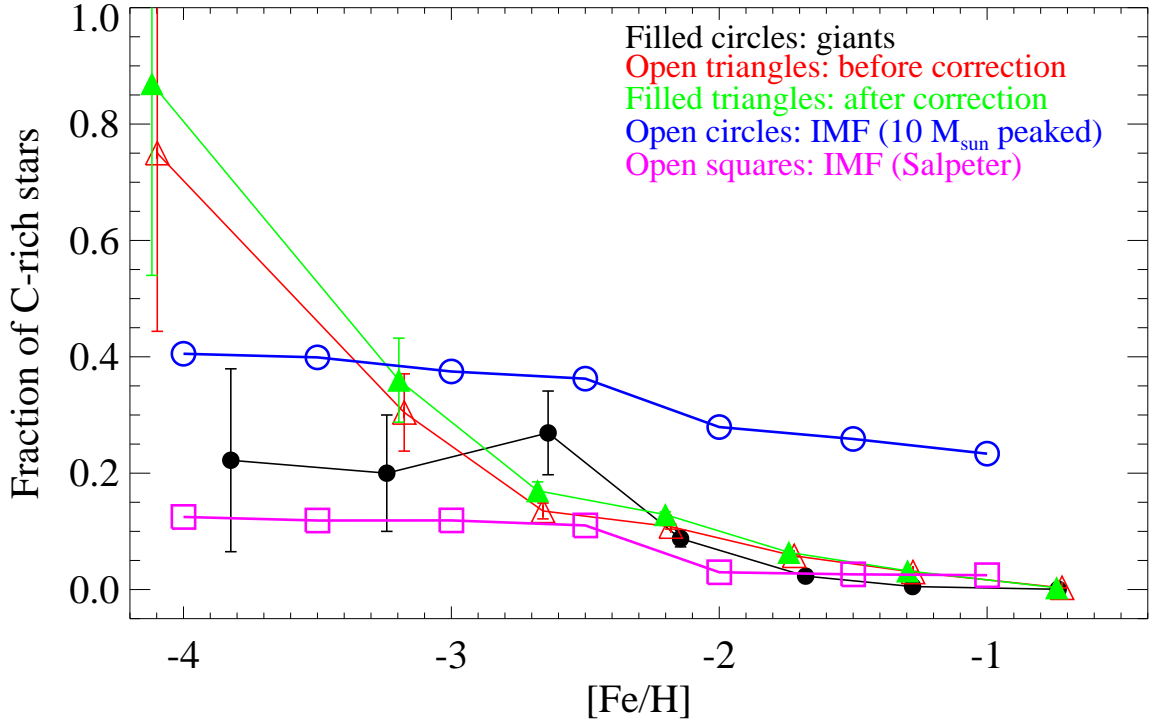}
\caption{Differential frequencies of C-rich main-sequence turnoff stars, as a function of
[Fe/H]. The metallicity bins are the same as in Figure
\ref{fig:cempcom}. Only stars located at $|Z| < $ 5 kpc are considered.
The red open triangles represent the ``as-observed'' frequencies for the
subsample of turnoff stars, defined as stars with parameters in the
range 5600~K $\leq$ \teff\ $\leq$ 6600~K and 3.2 $\leq$ \logg\ $\leq$
4.5, while the green filled triangles indicate the corrected CEMP
frequencies for this same sample (see text). The open circles and squares
are the AGB binary-synthesis model predictions for the two different
IMFs. For comparison, the frequencies of the CEMP giants with $|Z| <$ 5 kpc
are plotted as filled circles.}
\label{fig:cempcor}
\end{figure*}

In order to address this difficulty, and to provide a check on just how
many C-rich halo stars may have been misclassified as C-normal objects (\cfe\ $<
+0.7$) for stars around the turnoff region, we have performed the
following experiment. Following the prescription by Lee et al. (2013),
we inject artificial noise (with characteristics similar to that for a
typical SDSS/SEGUE spectrum) into the grid of synthetic spectra that are
used to estimate [C/Fe]. The noise-added synthetic spectra have S/N =
40, 45, and 50, which are typical of the quality of the SDSS/SEGUE
spectra in our study with $|Z| < 5$ kpc (justification for this
choice is provided below). At each S/N there are 25 different realizations.
These spectra are processed through the SSPP to determine estimates of [C/Fe]. With
the measured [C/Fe] in hand, we then derive the CEMP frequencies of the
spectra, as a function of [Fe/H], which have an {\it estimated} [C/Fe]
less than $+0.7$ among the spectra with $+0.75 \leq$ [C/Fe] $\leq +1.25$
for giants, turnoff stars, and dwarf stars (note that we employ discrete
[C/Fe] values, $+0.75, +1.0$, and $+1.25$ for the synthetic spectra).
For the purpose of this exercise, we define giants as models with
parameters in the ranges 4500~K $\leq$ \teff\ $\leq$~5500~K and 1.0
$\leq$ \logg\ $\leq 3.0$, turnoff stars for 5750~K $\leq$ \teff\ $\leq$
6500~K and 3.5 $\leq$ \logg\ $\leq 4.0$, and dwarfs for 4500~K $\leq$
\teff\ $\leq$ 5500~K and 4.5 $\leq$ \logg\ $\leq 5.0$.

Figure \ref{fig:corfun} shows the results of this experiment. The plus
signs represent the actual frequencies of CEMP stars, which are set to
1.0. The green squares indicate giants, the red circles the turnoff
stars, and the blue triangles the dwarfs. The bottom plot exhibits the
residuals in the derived fractions (simply 1 -- the fractions) in each
category. Inspection of this figure reveals that some of the C-rich
dwarfs and giants start to be classified as C-normal ([C/Fe] $<+0.7$)
from around [Fe/H] $< -2.0$; the number of the misclassified stars
slowly increases with decreasing metallicity. By way of comparison, the
C-rich turnoff stars begin to be misclassified as C-normal stars as
metallicity drops below [Fe/H] = $-1.0$; this misclassification rapidly
increases with declining metallicity, as expected. 

We now derive a correction function for capturing the ``true'' CEMP
frequency, as a function of [Fe/H], for the SDSS/SEGUE turnoff stars,
based on the results of the test carried out above. We then use this
correction function to adjust the frequency calculation, among the stars
with $+0.7 \leq$ \cfe\ $<+1.5$, taking the ``missing'' C-rich stars into
account. 

\subsection{Comparison with Model Predictions for the Frequencies of
C-rich Turnoff Stars}

Since the giants are more luminous than the turnoff stars, they can
probe to greater distances in the Galaxy. This increases the likelihood
of introducing a greater number of giants than turnoff stars into a
magnitude-limited sample (the frequency of giants can also be influenced
by the luminosity function of the halo field stars, as well as by 
shifts in the mix of stellar populations between the nearby and more
distant halo stars). This possible population transition has already
been noted in Figure \ref{fig:cempcom}, and discussed in detail in the
previous section. Thus, in order to make sure we are sampling the giants and
turnoff stars in similar regions of the Galaxy, we restrict our
considerations to the stars with $|Z| < 5$ kpc for the calculation of
the CEMP frequencies.

Figure~\ref{fig:cempcor} shows the differential frequencies of C-rich
stars for main-sequence turnoff stars, as a function of [Fe/H], as red
open triangles; the corrected frequencies are indicated as green filled
triangles. The blue open circles represent the predicted turnoff CEMP
frequencies from an AGB binary-synthesis model with an IMF peaked at 10
\msun, while the magenta open squares indicates the prediction obtained
using a Salpeter IMF. For comparison, the filled black circles are the
frequencies from the giants with $|Z| <$ 5 kpc given in
Figure~\ref{fig:cempcom}. The behavior seen in this figure is consistent
with our expectations, in that the CEMP frequencies for the giants are
lower than that for the turnoff sample, at least at the lowest
metallicities. Note that the corrected CEMP frequencies for the turnoff
stars are, on average, higher than the uncorrected frequencies, by about
5\%. The figure also shows that the models produce higher CEMP
frequencies for the turnoff stars than for the giants for both IMFs
(compare with Figure~\ref{fig:cempcom}), which at least qualitatively
agrees with the observations. The small difference in the
model-predicted CEMP frequencies between the giants and turnoff stars
(Figures~\ref{fig:cempcom} and \ref{fig:cempcor}) may arise from the
difference in the mass of the convective envelope; a star that evolves
to the giant stage has a much deeper convective zone, and hence more
effective dilution, resulting in lower CEMP frequencies derived for the giants.

Our derived CEMP frequencies from the distance-restricted turnoff sample
appears to be in good agreement with the model prediction based on a
Salpeter IMF (magenta open squares), down to \feh\ = $-2.5$. This
reaffirms that the AGB model for progenitor stars in the low-mass range
works well. However, the model estimate of the CEMP frequency does not
reproduce the observed frequencies for \feh\ $< -2.5$. Unlike the case
for the giants (Figure \ref{fig:cempcom}), the observed CEMP frequencies
from our turnoff sample do not agree with the model estimation from the
top-heavy IMF for \feh\ $< -2.5$ at all, as these remain roughly flat
instead of growing dramatically with decreasing metallicity.

One reason for the large discrepancy between the model estimates
and the observed frequencies of CEMP stars may be uncertainties of the
model parameters adopted for producing carbon in the AGB star, and 
subsequent processes that enrich (or deplete) the envelope with carbon.
Below we discuss known sources of uncertainty associated with the AGB
binary-synthesis model, which may result in changes of the
predicted carbon abundance of the secondary star. 

Most AGB stars in the mass range of $\sim$1--8 \msun\ can produce
carbon, but whether or not they develop carbon-enriched envelopes
depends on the efficiency of the third dredge-up (TDU) and helium-flash
driven deep mixing (He-FDDM; Fujimoto et al. 1990, 2000)\footnote[8]{Often referred to 
as a dual shell flash or carbon ingestion episode.} events for 
[Fe/H] $< -2.5$. At present, it is not fully understand how
such episodes depend on the mass and metallicity of an AGB star. 

For example, Lau et al. (2009) claimed, in a study of the evolution of AGB
stars with metallicity between Z = 10$^{-8}$ and 10$^{-4}$, that the
He-FDDM did not take place for Z $> 10^{-5}$ independent of the mass of
a star, and that this event did not occur for a star with $M > 2$
\msun\ , regardless of its metallicity. However, Suda \& Fujimoto (2010)
found, from an extensive set of stellar evolution models, that the He-FDDM
event occurred for a star with $M \leq 3$ \msun\ for zero metallicity,
while it occurred for a star with $M \leq 2$ \msun\ for $-5 \leq$ [Fe/H]
$\leq -3$. They also found that the TDU episode was restricted to a
mass range of $M \sim$ 1.5--5 \msun\ for [Fe/H] = $-3.0$, and that this
mass range becomes smaller as the metallicity decreases.

In the adopted models from Suda et al. (2013), the increasing fractions
of CEMP stars comes from the He-FDDM, which can enhance the surface
carbon abundance by a factor of 10, from [C/H] $\sim -1$ to $+0$,
regardless of the initial metallicity of the models, for stars with
masses of 0.8--3 \msun. In this view, the value of [C/Fe] for the
secondary component (the presently observed CEMP star) of a given binary
increases with decreasing [Fe/H], so that a larger fraction of CEMP
stars can be achieved at lower metallicity. It is also assumed that the
efficiency of the binary mass transfer and the mass-loss rates do not depend
on metallicity.

In addition, among these AGB stars, the intermediate-mass ($\sim$3--8 \msun) 
objects can be enriched with nitrogen by operation of the
hot-bottom burning (HBB) process, which converts carbon into nitrogen by
CN processing, and predicts the production of NEMP stars (Johnson et al.
2007). However, the dependency of the HBB on the mass and metallicity of
an AGB star is yet not well-established.

Even taking into account the uncertainties in the parameters of the AGB
models, a more plausible interpretation of our results may be the
existence of additional (non-AGB) carbon-production mechanisms, as
discussed in the Introduction, which result in large frequencies of CEMP
stars in the metallicity regime [Fe/H] $< -3.0$. The current
observations certainly favor this interpretation, since most CEMP-no
stars in the Galaxy appear at \feh\ $<-3.0$, and these stars do not
commonly exhibit the radial velocity variations that would be expected
if membership in a binary system were required (as in the AGB
mass-transfer scenario).

The adopted models, however, assume that all CEMP-no stars form from the AGB
binary mass-transfer scenario, rather than including additional sources that have
been argued are likely to be present in the early universe. According to
the AGB models, low-mass ($M < 3.5$ \msun) AGB stars efficiently create
$s$-process elements by generating extra neutrons via the
$^{13}$C($\alpha$,n)$^{16}$O reaction, while a weak $s$-process (for
light $s$-process elements) operates by $^{22}$Ne($\alpha$,n)$^{25}$Mg
for the intermediate-mass stars. CEMP-no stars could form in the AGB
mass-transfer scenario by suppressing the formation of the $^{13}$C
pocket for intermediate mass stars ($M > 3.5$ \msun; Suda et al. 2013).
Thus, in order to preferentially produce CEMP-no stars at low
metallicity and maintain the observed ratio of CEMP-no/(CEMP-$s$ $+$
CEMP-no), which is close to 0.5 at [Fe/H] $\sim -3.0$ (from Table 1 of
Suda et al. 2013), the models have to assume that the $^{13}$C pocket
does not form in stars with metallicity significantly below [Fe/H] =
$-2.5$. 

It has also been suggested (Komiya et al. 2007) that a secondary star
with an AGB primary having 0.8 \msun\ $< M < 3.5$ \msun\ could become a
CEMP-$s$ star following mass transfer, while systems that include an AGB
primary with $M > 3.5$ \msun\ could produce C without the enhancement of
neutron-capture elements, leading to a CEMP-no star following mass
transfer. However, current AGB models do not satisfactorily explain the
absence of the $^{13}$C pocket at low metallicities, even though this
assumption has been invoked to explain the observed decreasing trend of
[Pb/Ba] for CEMP stars with [Fe/H] $< -2.5$ (Aoki et al. 2002; Suda et
al. 2004; Barbuy et al. 2005; Cohen et al. 2006; Aoki et al. 2008). It
is also significant that Ito et al. (2013) obtained a rather low upper
limit on the abundance of lead (log $\epsilon$(Pb) $< -0.10$) for the
[Fe/H] $= -3.8$ CEMP-no star BD$+$44$^{\circ}$493, while previous
predictions called for log $\epsilon$(Pb) $\sim +1.5$ at these low
metallicities if the lead were produced by the $s$-process (Cohen et
al. 2006).

Chemical abundances of CEMP-no stars observed with high-resolution
spectroscopy (e.g., Ito et al. 2009, 2013; Norris et al. 2013) support
other scenarios for significant carbon production. Their abundance
patterns are similar to the predictions from massive, rapidly rotating,
MMP stars (Meynet et al. 2006, 2010) or faint SNe that
experience mixing and fallback (Umeda \& Nomoto 2003, 2005; Tominaga et
al.2007, 2013; Kobayashi et al. 2011; Ito et al. 2013; Nomoto et al. 2013). 
If such mechanisms are the dominant sources of the
large amounts of carbon produced at low metallicity, these scenarios also favor
an IMF that preferentially produces massive stars. In fact, one might
also expect a rather abrupt ``break'' in the CEMP frequencies when the
primary carbon sources change in nature---such a sudden change can be
seen for the turnoff stars in Figure \ref{fig:cempcor} below [Fe/H]
$= -3.0$. Although we are not able to distinguish CEMP-no stars from
CEMP-$s$ stars in our sample, given that the majority of the CEMP-no
stars are found with \feh\ $<-3.0$, our derived frequencies imply that
non-AGB related phenomenon may be the dominant mechanisms for producing
large carbon abundances at extremely low [Fe/H].

Finally, we see that a similar behavior in the CEMP frequencies with
[Fe/H] applies to the turnoff stars with $-2.5 <$ [Fe/H] $< -1.5$ as for
the giants, suggesting that the proposed shift in an IMF occurred over
this chemical interval.

\section{Conclusions}

We have compared our derived CEMP frequencies from the SDSS/SEGUE giant
sample with that predicted by AGB binary-synthesis models with two different
IMFs---a Salpeter IMF, and an IMF with a characteristic mass of 10
\msun. Good agreement of the CEMP frequencies for [Fe/H] $> -1.5$ with the
Salpeter IMF indicates that the adopted AGB model works well for
low-mass progenitor stars. Qualitatively, better agreement with an IMF
biased to higher-mass progenitors is found for [Fe/H] $<-2.5$,
suggesting that the nature of the IMF shifted from one that is high-mass 
dominated in the early history of the Milky Way galaxy, to one that is now
low-mass dominated. This transition appears to have occurred, in
``chemical time'', between [Fe/H] $= -2.5$ and [Fe/H] $= -1.5$, as other
recent studies have argued (e.g., Suda et al. 2011, 2013; Yamada et al.
2013).

As noted by other recent work, the more distant halo giants (those with
$|Z|$ $> 5$ kpc) exhibit higher frequencies of CEMP stars compared to
those closer to the Galactic plane. A plausible explanation for this
difference is the expected change of the dominant stellar populations from
the inner-halo to the outer-halo population, coupled with the assumption 
that the outer-halo stars are associated with progenitors capable of
producing large amounts of carbon without the accompanying production of
heavy metals.  Thus, one might expect that the inner-halo population
harbors a higher ratio of CEMP-$s$/CEMP-no stars, while the opposite may
apply to the outer-halo population. Tests of this hypothesis are
underway (D. Carollo et al., in preparation).   

The weak CH $G$-bands for moderately carbon-enhanced stars ($+0.7 <$
\cfe\ $<+1.5$) among warm, metal-poor main-sequence turnoff stars
results in their likely having been undercounted by previous assessments
of CEMP frequencies. We have derived a correction function to compensate
for this, making use of noise-added synthetic spectra. The corrected
CEMP frequencies for turnoff stars are, on average, higher by $\sim 5$\%
than with the uncorrected frequencies. Both the corrected and uncorrected
CEMP frequencies derived from the turnoff sample exceed those of the
giants for [Fe/H] $< -3.0$.

We have made use of main-sequence turnoff stars with $|Z| < 5$ kpc to
compute more realistic CEMP frequencies than obtained by using giants
(or the combination of giants with other classes), corrected as
mentioned above. For \feh\ $> -2.5$, our corrected CEMP frequencies
agree with the model predictions based on a Salpeter IMF, indicating
that the AGB model used in this study is probably not far from reality,
at least as applied to low-mass stellar progenitors. However, unlike the
case for the giant sample, the top-heavy IMF model does not reproduce
the observed trend of the CEMP frequencies for the turnoff stars at all.
The combination of these results from the giant and turnoff samples
suggests that the current AGB binary-synthesis model may not be suitable
for creating carbon-enhanced envelopes for intermediate- to high-mass
stars (3--8 \msun). As the AGB binary-synthesis model (using a
Salpeter IMF or a top-heavy IMF) predict far too low frequencies of CEMP
stars for our turnoff sample, there likely exists one or more additional
mechanisms capable of producing carbon-rich stars below [Fe/H] = $-3.0$,
the metallicity regime where the CEMP-no stars dominate over the
subclass of CEMP-$s$ stars.

\acknowledgments
Funding for SDSS-III has been provided by the Alfred P. Sloan Foundation, the 
Participating Institutions, the National Science Foundation, and the U.S. 
Department of Energy Office of Science. The SDSS-III Web site is 
http://www.sdss3.org/.

SDSS-III is managed by the Astrophysical Research Consortium for the 
Participating Institutions of the SDSS-III Collaboration including 
the University of Arizona, the Brazilian Participation Group, 
Brookhaven National Laboratory, University of Cambridge, Carnegie 
Mellon University, University of Florida, the French Participation 
Group, the German Participation Group, Harvard University, the 
Instituto de Astrofisica de Canarias, the Michigan State/Notre 
Dame/JINA Participation Group, Johns Hopkins University, Lawrence 
Berkeley National Laboratory, Max Planck Institute for Astrophysics, 
Max Planck Institute for Extraterrestrial Physics, New Mexico State 
University, New York University, Ohio State University, Pennsylvania 
State University, University of Portsmouth, Princeton University, 
the Spanish Participation Group, University of Tokyo, University of 
Utah, Vanderbilt University, University of Virginia, University of 
Washington, and Yale University. 

Y.S.L. is a Tombaugh Fellow. T.S. was supported by the JSPS Grants-in-Aid 
for Scientific Research (23224004). This work was supported in part 
by grant PHY 08-22648: Physics Frontiers Center/Joint Institute
for Nuclear Astrophysics (JINA), awarded by the U.S. National Science
Foundation.


\begin{thebibliography}{}
\bibitem[]{} Abazajian, K., Adelman-McCarthy, J. K., Ag\"ueros, M. A., et al. 2003, \aj, 126, 2081
\bibitem[]{} Abazajian, K., Adelman-McCarthy, J. K., Ag\"ueros, M. A., et al. 2004, \aj, 128, 502
\bibitem[]{} Abazajian, K., Adelman-McCarthy, J. K., Ag\"ueros, M. A., et al. 2005, \aj, 129, 1755
\bibitem[]{} Abazajian, K., Adelman-McCarthy, J. K., Ag\"ueros, M. A., et al. 2009, \apjs, 182, 543
\bibitem[]{} Abia, C., Busso, M., Gallino, R., et al. 2001, \apj, 559, 1117
\bibitem[]{} Adelman-McCarthy, J. K., Ag\"ueros, M. A., Allam, S. S., et al. 2006, \apjs, 162, 38
\bibitem[]{} Adelman-McCarthy, J. K., Ag\"ueros, M. A., Allam, S. S., et al. 2007, \apjs, 172, 634
\bibitem[]{} Adelman-McCarthy, J. K., Ag\"ueros, M. A., Allam, S. S., et al. 2008, \apjs, 175, 297
\bibitem[]{} Ahn, C. P., Alexandroff, R., Allende Prieto, C., et al. 2012, \apjs, 203, 21
\bibitem[]{} Aihara, H., Allende Prieto, C., An, D., et al. 2011, \apjs, 193, 29
\bibitem[]{} Allende Prieto, C., Sivarani, T., Beers, T. C., et al. 2008, \aj, 136, 2070
\bibitem[]{} Aoki, W., Beers, T. C., Christlieb, N., et al. 2007, \apj, 655, 492
\bibitem[]{} Aoki, W., Beers, T. C., Lee, Y. S., et al. 2013, \aj, 145, 13
\bibitem[]{} Aoki, W., Beers, T. C., Sivarani, T., et al. 2008, \apj, 678, 1351
\bibitem[]{} Aoki, W., Norris, J. E., Ryan, S. G., Beers, T. C., \& Ando, H. 2002, \apj, 580, 1149
\bibitem[]{} Barbuy, B., Spite, M, Spite, F., et al. 2005, \aap, 429, 1031
\bibitem[]{} Beers, T. C., Carollo, D., Ivezi\'c, \v Z., et al. 2012, \apj, 746, 34
\bibitem[]{} Beers, T. C., Chiba, M., Yoshii, Y., et al. 2000, \aj, 119, 2866
\bibitem[]{} Beers, T. C., \& Christlieb, N. 2005, \araa, 43, 531
\bibitem[]{} Beers, T. C., Preston, G. W., \& Shectman, S. A. 1985, \aj, 90, 2089
\bibitem[]{} Beers, T. C., Preston, G. W., \& Shectman, S. A. 1992, \aj, 103, 1987
\bibitem[]{} Bisterzo, S., Gallino, R., Straniero, O., Cristallo, S., \& K\"appeler, F. 2011, \mnras, 418, 284
\bibitem[]{} Bisterzo, S., Gallino, R., Straniero, O., Cristallo, S., \& K\"appeler, F. 2012, \mnras, 422, 849
\bibitem[]{} Bonifacio, P., Spite, M., Cayrel, R., et al. 2009, \aap, 501, 519
\bibitem[]{} Caffau, E., Bonifacio, P., Fran\c{c}ois, P., et al. 2011, \nat, 477, 67
%\bibitem[]{} Campbell, S. W., Lugaro, M., \& Karakas, A. I. 2010, \aap, 522, L6
\bibitem[]{} Carollo, D., Beers, T. C., Bovy, J., et al. 2012, \apj, 744, 195
\bibitem[]{} Chiappini, C. 2013, AN, 334, 5951
\bibitem[]{} Christlieb, N. 2003, RvMA, 16, 191
\bibitem[]{} Christlieb, N., Bessell, M. S., Beers, T. C., et al. 2002, \nat, 419, 904
\bibitem[]{} Christlieb, N., Green, P. J., Wisotzki, L., \& Reimers, D. 2001, \aap, 375, 66
\bibitem[]{} Christlieb, N., Sch\"orck, T., Frebel, A., et al. 2008, \aap, 484, 721   
\bibitem[]{} Cohen, J. McWilliam, A., Shectman, S., et al. 2006, \aj, 132, 137   
%\bibitem[]{} Cruz, M. A., Serenelli, A., \& Weiss, A. 2013, \aap, in press, arXiv:1308.2224   
\bibitem[]{} Frebel, A., Christlieb, N., Norris, J. E., et al. 2006, \apj, 652, 1585
\bibitem[]{} Fujimoto, M. Y., Iben, I. Jr., \& Hollowell, D. 1990, \apj, 349, 580
\bibitem[]{} Fujimoto, M. Y., Ikeda, Y. \& Iben, I. Jr. 2000, \apj, 529, L25
\bibitem[]{} Fukugita, M., Ichikawa, T., Gunn, J. E., et al. 1996, \aj, 111, 1748
\bibitem[]{} Gunn, J. E., Carr, M., Rockosi, C., et al. 1998, \aj, 116, 3040
\bibitem[]{} Gunn, J. E., Siegmund, W. A., Mannery, E. J., et al. 2006, \aj, 131, 2332
\bibitem[]{} Hansen, T., Andersen, J., \& Nordtro\"m, B. 2013, XII International Symposium on Nuclei 
             in the Cosmos, arXiv:1301.7208
\bibitem[]{} Herwig, F. 2005, \araa, 43, 435
\bibitem[]{} Ito, H., Aoki, W., Beers, T. C., et al. 2013, \apj, 773, 33
\bibitem[]{} Ito, H., Aoki, W., Honda, S., \& Beers, T. C. 2009, \apj, 698, L37
\bibitem[]{} Izzard, R. G., Glebbeek, E., Stancliffe, R. J., \& Pols, O. R. 2009, \aap, 508, 1359
\bibitem[]{} Johnson, J. A., Herwig, F., Beers, T. C., \& Christlieb, N. 2007, \apj, 658, 1203
\bibitem[]{} Kobayashi, C., Tominaga, N., \& Nomoto, K. 2011, \apj, 730, L14
\bibitem[]{} Komiya, Y., Suda, T., Minaguchi, H., et al. 2007, \apj, 658, 367
\bibitem[]{} Korn, A., Grundahl, F., Richard, O., et al. 2007, \apj, 671, 402
\bibitem[]{} Lau, H. H. B., Stancliffe, R. J., \& Tout, C. A. 2009, \mnras, 396, 1046
\bibitem[]{} Lee, Y. S., Beers, T. C., Allende Prieto, C., et al. 2011, \aj, 141, 90
\bibitem[]{} Lee, Y. S., Beers, T. C., Masseron, T., et al. 2013, \aj, in press
\bibitem[]{} Lee, Y. S., Beers, T. C., Sivarani, T., et al. 2008a, \aj, 136, 2022
\bibitem[]{} Lee, Y. S., Beers, T. C., Sivarani, T., et al. 2008b, \aj, 136, 2050
\bibitem[]{} Lind, K., Korn, A. J., Barklem, P. S., \& Grundahl, F. 2008, \aap, 490, 777
\bibitem[]{} Lucatello, S., Beers, T. C., Christlieb, N. C., et al. 2006, \apj, 652, L37
\bibitem[]{} Lucatello, S., Gratton, R. G., Beers, T. C., \& Carretta, E. 2005b, \apj, 625, 833
\bibitem[]{} Lucatello, S., Tsangarides, S., Beers, T. C., et al. 2005a, \apj, 625, 825
\bibitem[]{} Luck, R. E., \& Bond, H. E. 1991, \apjs, 77, 515
\bibitem[]{} Marsteller, B., Beers, T. C., Rossi, S., et al. 2005, Nucl. Phys. A., 758, 312
\bibitem[]{} Masseron, T., Johnson, J. A., Plez, B., et al. 2010, \aap, 509, 93
\bibitem[]{} Meynet, G., Ekstr\"om, S., \& Maeder, A.. 2006, \aap, 447, 623
\bibitem[]{} Meynet, G., Hirschi, R., Ekstrom, S., et al. 2010, \aap, 521, 30
\bibitem[]{} Nomoto, K., Kobayashi, C., \& Tominaga, N. 2013, \araa, 51, 457
\bibitem[]{} Norris, J.E., Ryan, S.G., \& Beers, T.C. 1997, \apj, 488, 350
\bibitem[]{} Norris, J. E., Christlieb, N., Korn, A. J., et al. 2007, \apj, 670, 774
\bibitem[]{} Norris, J. E., Yong, D., Bessell, M. S., et al. 2013, \apj, 762, 28
\bibitem[]{} Pier, J.R., Munn, J. A., Hindsley, R. B., et al. 2003, \aj, 125, 1559
\bibitem[]{} Pols, O. R., Izzard, R. G., Stancliffe, R. J., \& Glebbeek, E. 2012, \aap, 547, 76
\bibitem[]{} Richard, O., Michaud, G., \& Richer, J. 2002a, \apj, 580, 1100
\bibitem[]{} Richard, O., Michaud, G., Richer, J., et al. 2002b, \apj, 568, 979
\bibitem[]{} Roederer, I. U. 2009, \aj, 137, 272
\bibitem[]{} Rossi, S., Beers, T.C., \& Sneden, C. 1999, in The Third Stromlo Symposium, ASP Conference 
             Series, eds. B.K. Gibson, T.S. Axelrod, and M.E. Putman, 165, p. 264
\bibitem[]{} Rossi, S., Beers, T. C., Sneden, C., et al. 2005, \aj, 130, 2804
\bibitem[]{} Smolinski, J. P., Lee, Y. S., Beers, T. C., et al. 2011, \aj, 141, 89
\bibitem[]{} Sneden, C., Cowan, J. J., \& Gallino, R. 2008, \araa, 46, 241
\bibitem[]{} Spite, M., Caffau, E., Bonifacio, P., et al. 2013, \aap, 552, 107
\bibitem[]{} Spite, M., Cayrel, R., Hill, V., et al. 2006, \aap, 455, 291
\bibitem[]{} Spite, M., Cayrel, R., Plez, B., et al. 2005, \aap, 430, 655
\bibitem[]{} Stoughton, C., Lupton, R. H., Bernardi, M., et al. 2002, \aj, 123, 485
\bibitem[]{} Suda, T., Aikawa, M., Machida, M. N., \& Fujimoto, M. Y. 2004, \apj, 611, 476
\bibitem[]{} Suda, T., \& Fujimoto, M. Y. 2010, \mnras, 405, 177
\bibitem[]{} Suda, T., Katsuta, Y., Yamada, S., et al. 2008, PASJ, 60, 1159
\bibitem[]{} Suda, T., Komiya, Y., Yamada, S., et al. 2013, \mnras, 432, L46
\bibitem[]{} Suda, T., Yamada, S., Katsuta, Y., et al. 2011, \mnras, 412, 843
\bibitem[]{} Tominaga, N., Iwamoto, N., \& Nomoto, K. 2013, \apj, submitted, arXiv:1309.6734
\bibitem[]{} Tominaga, N., Umeda, H., \& Nomoto, K. 2007, \apj, 660, 516
\bibitem[]{} Tumlinson, J. 2007,\apj, 664, L63
\bibitem[]{} Umeda, H., \& Nomoto, K. 2003, \nat, 422, 871
\bibitem[]{} Umeda, H., \& Nomoto, K. 2005, \apj, 619, 427
\bibitem[]{} Wisotzki, L., Koehler, T., Groote, D., \& Reimers, D. 1996, A$\&$AS, 115, 227
\bibitem[]{} Wood P. R., 2011, in Qain S., Leung K., Zhu L., Kwok S., eds, ASP Conf.
             Ser. Vol. 451, Proc. 9th Pacific Rim Conference on Stellar Astrophysics 
             Astron. Soc. Pac., San Francisco, p. 87
\bibitem[]{} Yamada, S., Suda, T., Komiya, Y., Aoki, W., \& Fujimoto, M. Y. 2013, \mnras, in press, arXiv:1309.3430
\bibitem[]{} Yanny, B., Newberg, H. J., Johnson, J. A., et al. 2009, \aj, 137, 4377
\bibitem[]{} Yong, D., Norris, J. E., Bessell, M. S., et al. 2013, \apj, 762, 27
\bibitem[]{} York, D. G., Adelman, J., Anderson, J. E., Jr., et al. 2000, \aj, 120, 1579
\end{thebibliography}
\end{document}